\documentclass{elsart}

\usepackage{graphicx}

\usepackage{amssymb}

\newcommand{\lzero}{\mbox{Level-0}}

\newcommand{\pt}{\mbox{$p_{\rm T}$}}
\newcommand{\CP}{\mbox{$CP$}}
\newcommand{\x}{\mbox{$x$}}
\newcommand{\y}{\mbox{$y$}}
\newcommand{\z}{\mbox{$z$}}

\newcommand{\particle}[1]{{\ensuremath{\rm #1}}}

\renewcommand{\b}{\particle{b}}
\newcommand{\bbar}{\particle{\bar{b}}}
\newcommand{\bb}{\particle{b\bar{b}}}
\renewcommand{\e}{\particle{e^-}}

\newcommand{\pz}{\particle{\pi^0}}

\newcommand{\g}{\particle{\gamma}}

\begin{document}

\begin{frontmatter}



\title{The Level-0 Muon Trigger for the LHCb Experiment}


\author[cppm]{E. Aslanides}
\author[cppm]{J.-P. Cachemiche}
\author[cppm]{J. Cogan}
\author[cppm]{B. Dinkespiler}
\author[cppm]{S. Favard\thanksref{sf}}
\author[cppm]{P.-Y. Duval}
\author[cppm]{R. Le Gac\corauthref{rlg}}
\author[cppm]{O. Leroy}
\author[cppm]{P.-L. Liotard}
\author[cppm]{F. Marin}
\author[cppm]{M. Menouni}
\author[cppm]{A. Roche}
\author[cppm]{A. Tsaregorodtsev}

\address[cppm]{Centre de Physique des Particules de Marseille,\\
Aix--Marseille Universit\'e,CNRS/IN2P3, Marseille, France}

\thanks[sf]{Now at Observatoire de Haute Provence, Saint-Michel de l'Observatoire, France}
\corauth[rlg]{Corresponding author.}

\begin{abstract}
A very compact architecture has been developed for the first level Muon Trigger of the LHCb experiment
that processes $40 \times 10^6$ proton-proton collisions per second.
For each collision, it receives 3.2~kBytes of data and it finds straight tracks
within a 1.2~$\mu$s latency.
The trigger implementation is massively parallel, pipelined and
fully synchronous with the LHC clock.
It relies on 248 high density Field Programable Gate arrays
and on the massive use of multigigabit serial link transceivers embedded inside FPGAs.
\end{abstract}

\end{frontmatter}

\section{Introduction}\label{Introduction}
The LHCb experiment \cite{LHCbTDR} is installed at the Large Hadron Collider at CERN,
to study \CP\ violation and rare decays in the beauty sector.
Interesting b-hadron decays have to be isolated in a large background,
in proton-proton collisions at a center of mass energy of 14~TeV.
The cross-section for producing a \bb\ pair is around 500~$\mu$b while the inelastic
cross section is 80 mb.
In addition, branching ratios for interesting b-meson
final states vary between $10^{-3}$ and $10^{-9}$.

\begin{figure*}
  \centering
  \includegraphics*[width=\textwidth]{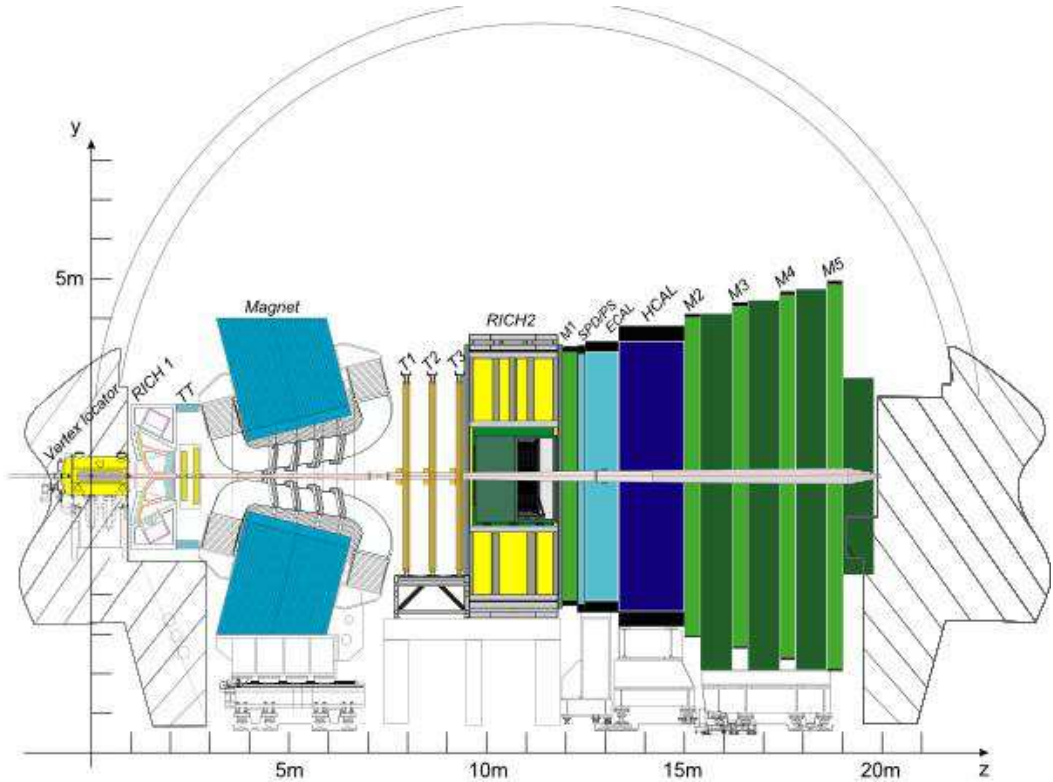}
  \caption{Vertical cross-section of the LHCb detector.}
\label{FIG_1}
\end{figure*}

The LHCb detector, shown in Figure~\ref{FIG_1}, is a single-arm spectrometer covering the forward
region of the proton-proton interactions.
The pseudo-rapidity domain ranges between 1.9 and 4.9.
This geometry is driven by the kinematics of the \bb\ pair production at the LHC energy where
both \b\ and \bbar\ quarks mainly go in the forward or backward direction.
The interaction point is surrounded by the Vertex locator,
a silicon strip $(r,\, \varphi)$ detector measuring precisely the position of primary and secondary vertices.
It also houses the pile-up detector which counts the number of interactions per collision.
Ring Imaging Cherenkov counters, RICH1 and RICH2, identify kaons and pions in the momentum
range of the experiment, 2--100~GeV/$c$.
A warm dipole magnet produces a vertical field with a bending power of 4 Tm
in the horizontal plane.
The tracking is performed by two groups of tracking stations located before and after the magnet,
the TT and T stations respectively.
They measure the kick given by the magnetic field, in order to determine the
momentum of the track with an accuracy of $4 \times 10^{-3}$.
A Scintillator pad detector, a preshower, an electromagnetic and a hadronic calorimeter identify \e , \g ,
hadrons and \pz\ and measure their energy.
The muon detector is composed of five stations sandwiched between iron shielding blocks.
The LHCb detector is designed to run at a luminosity of $2 \times 10^{32}$~cm$^{-2}$s$^{-1}$,
much lower than the nominal luminosity of the LHC machine. The luminosity for LHCb is
tunable locally whereas Atlas and CMS operate at the highest possible luminosity.

The LHCb trigger is divided in two systems: the \lzero\ trigger and the
High Level Trigger.
The purpose of the \lzero\ is to reduce the LHC beam crossing rate from 40~MHz to
1~MHz where the entire detector can be read out.
The \lzero\ is based on custom electronics collecting dedicated information from
the pile-up, calorimeters and muon detectors.
It looks for electrons, muons, \g s and hadrons with a high transverse energy due to the large mass
of b-hadrons.
The latency to process a proton-proton collision is limited to 4~$\mu$s.
This time includes the time-of-flight, cable length and all delays in the front-end
electronics, leaving 2~$\mu$s for the processing of the data in the \lzero\ trigger.
The purpose of the High Level Trigger is to reduce the rate down to 2~kHz by
using data from all sub-detectors.
It is based on a farm of a thousand of computers interconnected
through a gigabit Ethernet network.
It refines candidates found by the \lzero\ trigger
looking for tracks with a high transverse momentum and large impact parameter.
Then interesting final states are selected using inclusive and exclusive
criteria.
The measurements aimed for by LHCb require a very high
precision: hence systematic errors must be controlled to a very high degree. Amongst
the 2~kHz of accepted events, a large fraction is dedicated to precise calibration
and monitoring of the detector and its performance.

The \lzero\ trigger is subdivided in three components: the pile-up system,
the \lzero\ calorimeters and the \lzero\ muon. Each component is connected
to a dedicated detector and to the central \lzero\ decision unit collecting all candidates
to make the final decision.
The requirements for a \lzero\ subsystem are the following: input rate 40~MHz,
time to process data limited to a maximum of 2~$\mu$s;
fully synchronous with the LHC clock;
all collisions have to be processed.

This paper describes in detail the \lzero\ muon trigger.
The next section gives an overview of the Muon system.
Architecture and implementation of the trigger are described in sections 3 and 4.
Technologies used are discussed in section 5, debugging and monitoring tools
in section 6.
\section{Overview of the Muon System}\label{Overview}
The muon system \cite{LHCbTDR} has been designed to identify muons with a high transverse
momentum: a typical signature of a b-hadron decay. It is divided in two subsystems
intimately related: the muon detector and the \lzero\ muon trigger.
The system was optimized to perform an efficient muon identification and
standalone muon track reconstruction with a \pt\ resolution of 20\%.

\begin{figure*}
  \centering
  \includegraphics*[width=0.9\textwidth]{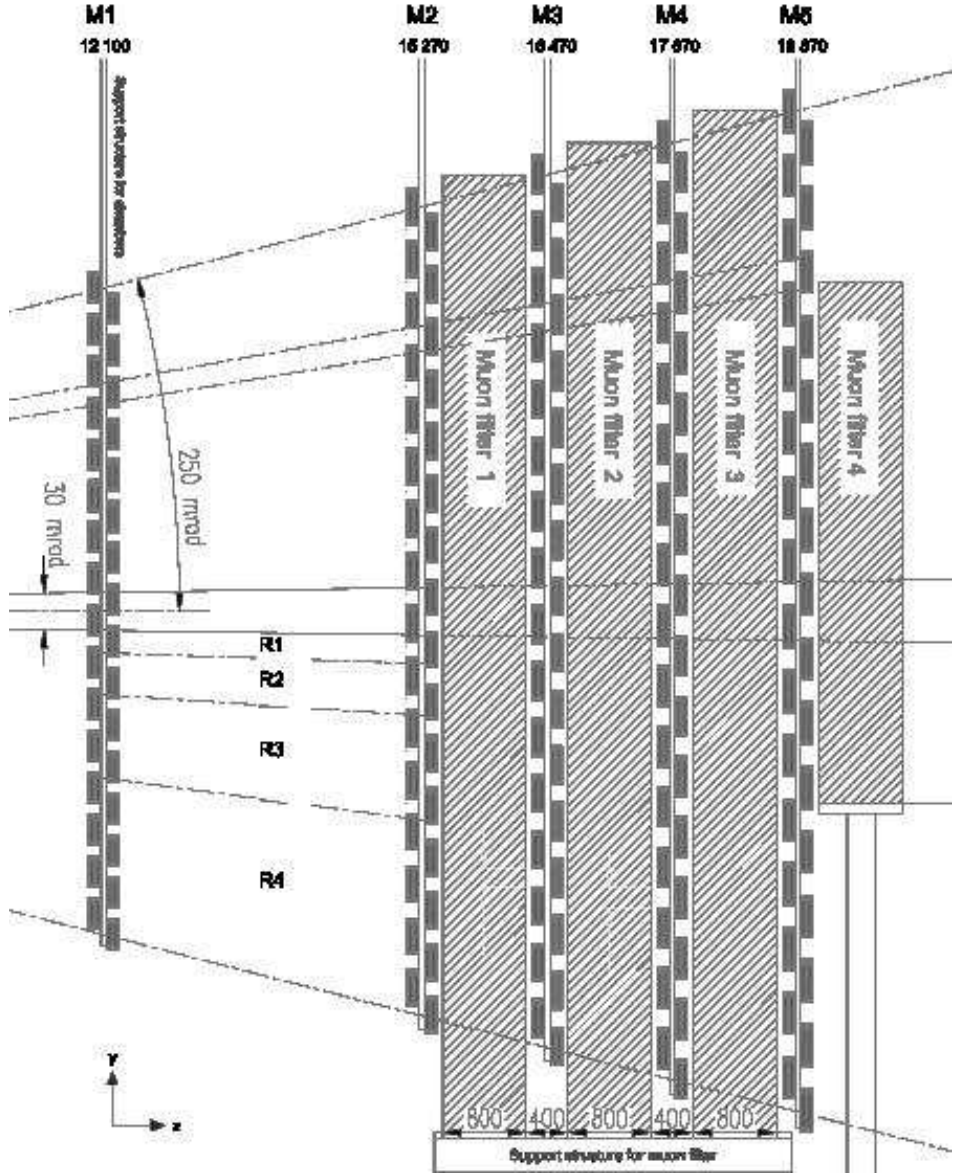}
  \caption{Side view of the muon system in the \y ,\z\ plane showing the iron filters,
  stations and chambers organization as well as the projectivity of the system.}
\label{FIG_2}
\end{figure*}

The muon detector, shown in Figure~\ref{FIG_2}, consists of five muon stations
M1-M5 interleaved with muon filters.
The first filter, between station M1 and M2, consists of the electromagnetic and
hadronic calorimeters. It is followed by four iron absorbers.
Stations provide binary space point measurements of tracks.
They are segmented in pads that are thinner in the magnet bending
plane to give an accurate measurement of the \pt .
The pad size depends on the station and on its location in the station.
Along the \x -axis, it is twice smaller for M2-M3 and twice coarser for
M4-M5 with respect to M1.
The segmentation is projective to ease the tracking in the \lzero\ muon trigger:
hence all stations cover the same angular acceptance and the pad size
scales with the distance from the interaction point.

The muon detectors are subjected to an intense flux of charged and neutral particles
varying between $\sim$45~Hz/cm$^2$ for the outer part of station M5
and $\sim$230~kHz/cm$^2$ for the inner part of station M1.
Multi-wire proportional chambers have been adopted for all stations.
However, triple-GEM (Gas Electron Multiplier) chambers are used for the inner part of station M1
where the rate is very high for a multi-wire proportional chamber.

The muon detector is composed of 1~368 multi-wire proportional chambers
and 12 GEM chambers.
The total surface covered by all chambers is equal to 435~m$^2$.
\subsection{Muon detector}
The first station M1 is placed in front of the calorimeter preshower at
about 12~m from the interaction point while the last station is at about 19~m.
The dimensions of stations M1 and M5 are $7.7\times6.4$~m$^2$ and $11.9\times9.9$~m$^2$
respectively. The size of logical pad varies between $0.5\times2.5$~cm$^2$ for the inner
part of station M2 and $16\times20$~cm$^2$ for the outer part of station M5.

Stations M2-M3 are devoted to the muon track finding while stations M4-M5
confirm the muon identification. The station M1 plays an important role for
the \pt\ measurement of the muon track improving its resolution by about 30\%.

The trigger algorithm uses a five-fold coincidence, the efficiency for each sation
must be at least 99\%, with a time resolution better than 25~ns in order to
unambiguously identify the bunch crossing.
Each station has two independent detector layers, logically OR-ed on the chamber, to
form {\em logical channels}. The independence of the detector layers provides a high
degree of redundancy.
The total number of logical channels is 25~920.

\begin{figure*}
  \centering
  \includegraphics*[width=\textwidth]{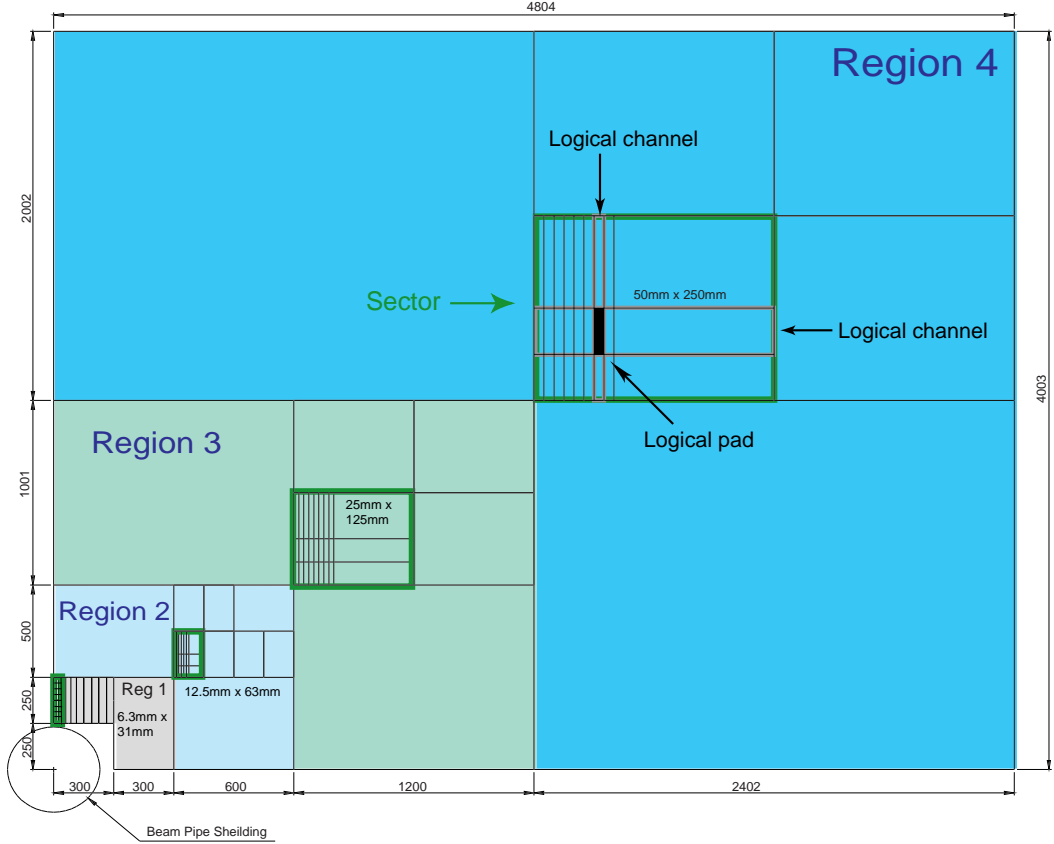}
  \caption{Front view of one quadrant of muon station M2, showing the dimension of the regions.
           Inside each region is shown a sector, defined by the size of the horizontal and
           vertical strips.}
\label{FIG_3}
\end{figure*}
Each station is subdivided into four regions with different {\em logical pad dimensions},
as shown in Figure~\ref{FIG_3}.
Regions and pad sizes scale by a factor two from one region to the next.
The logical layout in the five muon stations is projective in \y\ to the interaction point.
It is also projective in \x\ when the bending in the horizontal plane introduced
by the magnetic field is ignored.

Pads are obtained by the crossing of horizontal and vertical strips when applicable.
Strips are employed in stations M2-M5 while station M1 and region R1
of stations M4-M5 are equipped with pads.
Strips allow a reduction in the number of logical channels to be transferred
to the muon trigger. The processor receives 25~920 logical channels every 25~ns
corresponding to 55~296 logical pads obtained by crossing strips.

Each region is subdivided into {\em sectors} as shown in Figure~\ref{FIG_3}.
They are defined by the size of the horizontal and vertical strips and
match the dimension of underlying chambers.
\subsection{\lzero\ muon trigger}
\begin{figure*}
  \centering
  \includegraphics*[width=\textwidth]{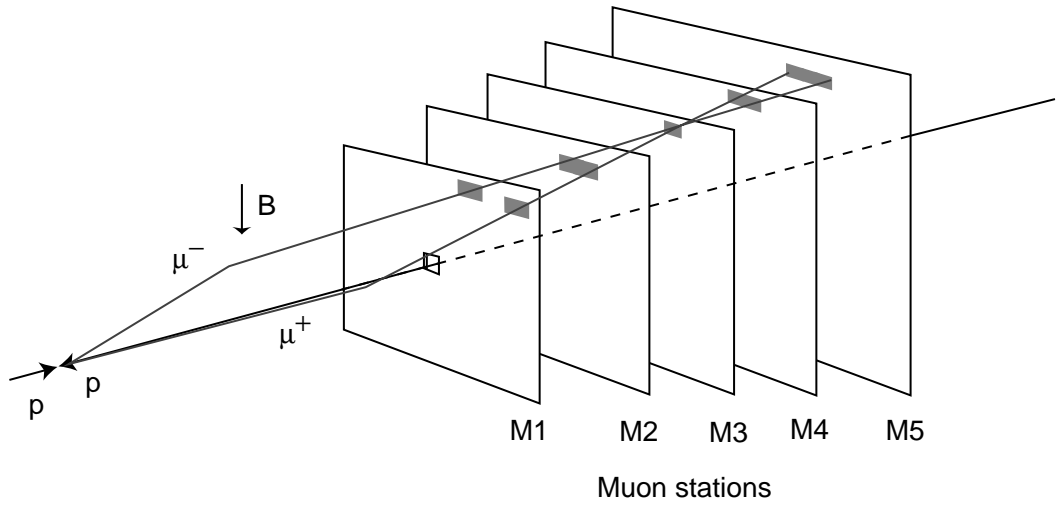}
  \caption{Track finding by the \lzero\ muon trigger. In the example shown,
           $\mu^+$ and $\mu^-$ cross the same pad in M3.
           Grey areas illustrate the field of interests used by the algorithm for station M1, M2, M4 and M5.}
\label{FIG_4}
\end{figure*}
The \lzero\ muon trigger looks for muon tracks with a large \pt .
The track finding is performed using the logical pad information.
It searches for hits defining a straight line
through the five muon stations and pointing towards the interaction point,
as shown in Figure~\ref{FIG_4}.
The position of a track in the first two stations allows the
determination of its \pt .

\begin{figure*}
  \centering
  \includegraphics*[width=\textwidth]{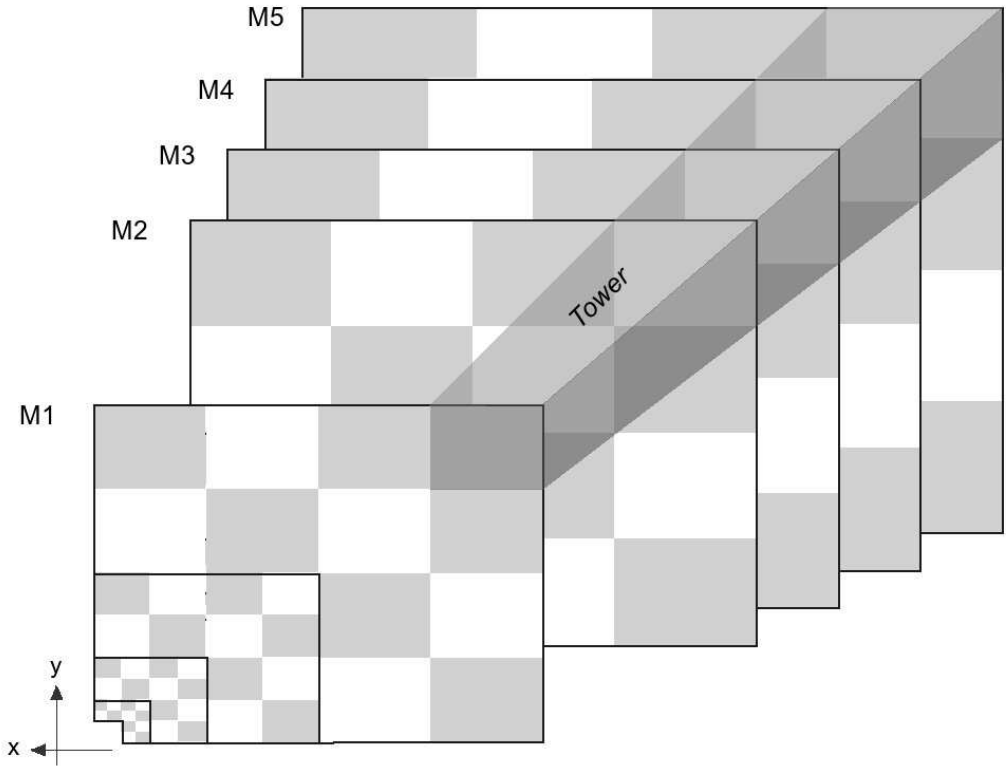}
  \caption{A quadrant of the muon detector with its 48 towers pointing toward
  the interaction point.}
\label{FIG_5}
\end{figure*}

To simplify the processing and to hide the complex layout of stations,
the muon detector is subdivided into $48\times4=192$ {\em towers} pointing to the
interaction point.
The tower organization is shown for a quadrant of the muon detector in Figure~\ref{FIG_5}.
All towers contain logical pads with 48 pads from M1,
96 pads from M2, 96 pads from M3, 24 pads from M4 and 24 pads from M5.
Therefore the same algorithm can be executed in all towers.
Each tower is  connected to a {\em processing unit}, the key component
of the trigger processor.
\subsection{Track finding algorithm}
\begin{figure*}
  \centering
  \includegraphics*[width=\textwidth]{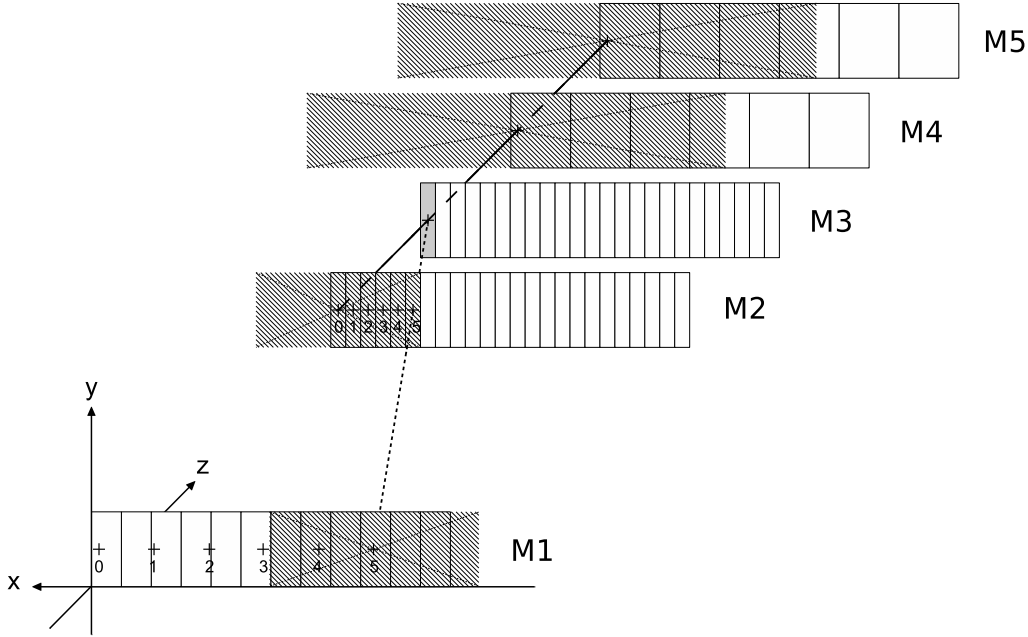}
  \caption{Fields of interest associated with a M3 pad located on the left border of a tower.
           The straight line shows the extrapolated position in stations M2, M4 and M5.
           The hashed arrays show the maximum size of the field of interest
           centered on the extrapolated position where hits are searched.
           The dash line shows the straight line extrapolation from M3 and M2 to M1
           when the pad labeled 5 is hit in M2.
           Numbers show the one-to-one correspondence between a pair of pads hit in M3-M2 and the
           extrapolated position in M1.
           The track finding is also performed along the \y -axis for station M4 and M5.
           The \y -size of the field of interest is $\pm1$ pad.
           It is not drawn for simplicity.}
\label{FIG_6}
\end{figure*}

The track finding is based on a road algorithm illustrated in Figures~\ref{FIG_4} and \ref{FIG_6}.
It assumes muon tracks coming from the interaction point with a single kick from the magnet.

For each logical pad hit in M3, the straight line passing through
the hit and the interaction point is extrapolated to M2, M4 and M5.  Hits are
looked for in these stations in search windows, called fields of interest,
approximately centred on the straight-line extrapolation. The size of the
field of interest depends on the muon station considered, the distance from
the beam axis, the level of background, and the minimum-bias retention
required.
When at least one hit is found inside the field of interest for
each of the stations M2, M4 and M5, a muon track is flagged and the pad hit
in M2 closest to the extrapolation from M3 is selected for subsequent use.

The track position in station M1 is determined by making a straight-line
extrapolation from M3 and M2, and identifying in the M1 field of interest
the pad hit closest to the extrapolation point.

Since the logical layout is projective, there is a one-to-one mapping from
pads in M3 to pads in M2, M4 and M5.  There is also a one-to-one mapping
from pairs of pads in M2 and M3 to pads in M1.  This allows the track-finding
algorithm to be implemented using only logical operations.

Once track finding is completed, a maximum of 96 candidates can be found,
one per M3 pad of the tower.
The two closest to the beam are selected and the remaining ones are dropped.
The \pt\ of the two selected tracks is determined from the track hits in
M1 and M2, using look-up tables.

The track finding is run on each quadrant of the muon system independently.
Therefore, the two muon tracks of highest \pt\ are selected for each quadrant
and the information for up to eight selected tracks is transmitted to the \lzero\
decision unit.

To satisfy realtime constraints, track finding algorithms are run in parallel
for each pad of the station M3 and for each proton-proton collision.
Therefore, the \lzero\ muon trigger executes
$192 \times 96 \times 40 \times 10^6 = 737\times 10^9$ algorithms per second.
\subsection{A complex data flow}
The implementation of the track finding algorithm is complex:
large number of logical channels distributed in a large volume;
mixture of pads and strips; segmentation of logical pads varying between
regions and stations;
one-to-one correspondence between towers and trigger sectors except for region R1
of stations M2-M3 where a trigger sector is shared by two towers and in region R2
where a tower maps two sectors (see Figure~\ref{FIG_3}).

Each processing unit gathers a large number of logical channels. It receives
an equivalent of 288 pads from its tower every 25~ns.
It also has to exchange a significant amount of data with its neighbours to avoid
inefficiency on the borders of the tower.
The quantity of logical channels is determined by the maximum width of the
fields of interest.
A processing unit emits 224 and receives 214 logical channels, in the worst case.

The granularity of neighbours is often different since fields of interest
are open along the \x -axis for stations M1-M2 and along the \x\ and \y -axis for
station M4-M5. A processing unit exchanges data with up to eleven neighbours since
the track finding works in a space with a uniform granularity.
In such configurations, the pattern of data exchange depends strongly on the
location of the tower.

Dedicated software tools have been developed to describe the topologies of exchange and to
store them in a database. The database is used by CAE tools, by the emulator and
by monitoring/debugging software.

\section{Architecture for a \lzero\ Muon processor}\label{Architecture}

\begin{figure*}
  \centering
  \includegraphics*[width=\textwidth]{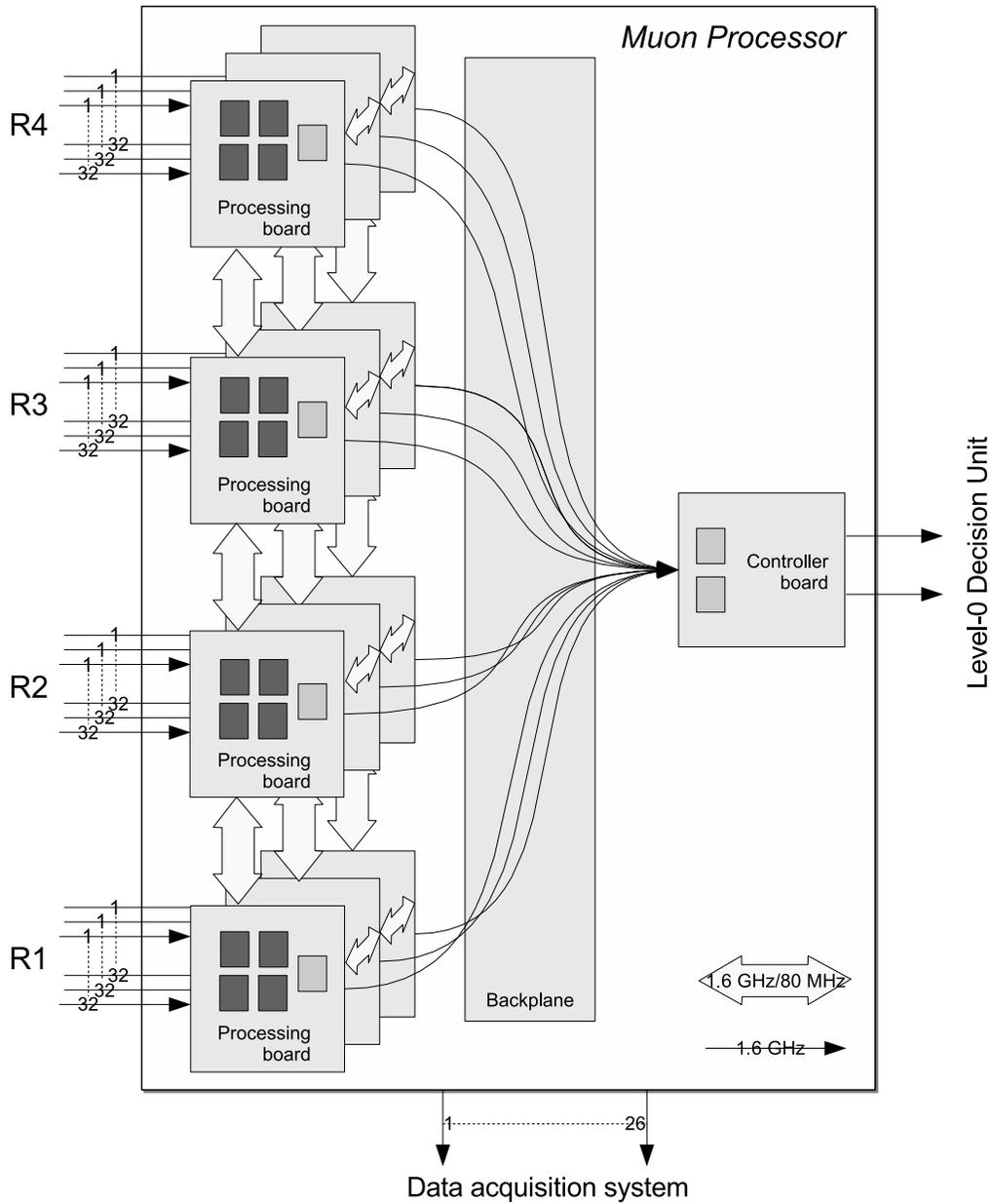}
  \caption{Overview of a \lzero\ muon processor.}
\label{FIG_7}
\end{figure*}

An overview of the \lzero\ architecture is given in Figure~\ref{FIG_7}.
Each quadrant of the muon detector is connected to a separate \lzero\ muon processor
through 312 optical links.
The latter are made of 36 ribbons containing 12 optical fibres each.
An optical link transmits a 32-bit word every 25~ns by serializing data at
1.6~Gbps.

A \lzero\ muon processor is a 9U crate containing
12 processing boards, one controller board and a custom backplane.
It is also connected to the \lzero\ decision unit which collects all muon candidates,
to the data acquisition system of the experiment and to the Timing Trigger and Control (TTC)
distribution system of LHC~\cite{TTC}.

A processing board houses five processing elements, four PUs (Processing Unit) and
one BCSU (Best Candidate Selection Unit). A PU runs 96 tracking algorithms in parallel,
one for each M3 pad of the tower, while the BCSU selects the two muons with the highest
momentum within the board.

A controller board houses a {\em control unit} and a {\em slave unit}.
They receive 24 candidates from 12 processing boards, select the two with the highest \pt\
and send them to the \lzero\ decision unit.
This board also distributes the 40~MHz clock and TTC signals via the crate backplane.

The custom backplane is necessary for the exchange of logical channels between PUs located
on different boards and to distribute the master clock and TTC signals.

All logical channels belonging to a tower are sent to a PU using a maximum of eight optical links:
two for station M1, one or two for M2, one or two for M3, one for M4 and one for M5.
Such an organization increases the number of links but eases the connectivity between
the muon detector and the trigger, avoiding a complex data distribution at the input
of a processing board.

The radiation level expected at the muon front-end electronics can reach
22~Gy after ten years of operation.
Optical links allow to place \lzero\ muon processors in the counting room,
far away from the detector, in a radiation free environment with full access.
The muon trigger is therefore immune to SEU (Single Event Upset) and
off the shelf components can be used.
However, the trigger interface, located in the the muon front-end electronics \cite{MFE},
is sensitive to SEU through the optical
drivers implemented in the interface (see section~\ref{OL}).

\begin{figure*}
  \centering
  \includegraphics*[width=\textwidth]{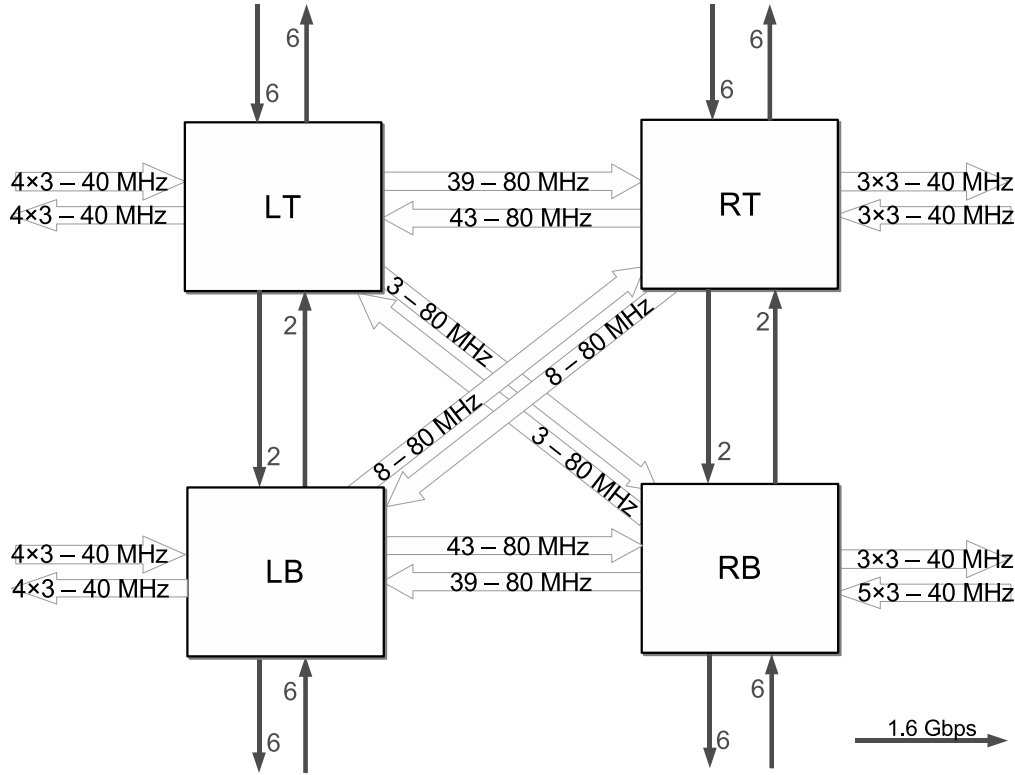}
  \caption{Interconnections between PUs located on a processing board. They are based on
  40~MHz parallel links, 80~MHz double data rate links and 1.6~Gbps serial links.
  Links on the periphery are used to connect PUs located on different boards. }
\label{FIG_8}
\end{figure*}

A single generic processing board was designed but configurations loaded in each PU
differ according to the area covered by the board.
The number of configurations is equal to 48, one per tower of a quadrant.
The size of each connection between PUs has been maximized to accommodate all configurations.
We use 40~MHz parallel links, 80~MHz double data rate links and 1.6~Gbps serial links.
The resulting topology of the data exchange is shown in Figure~\ref{FIG_8}.

The PUs are arranged following a $2\times2$ matrix.
The left and right columns are interconnected to allow the data exchange required
by the muon tracking along the \x -axis.
The top and  bottom rows are interconnected to allow the data exchange required
by the muon tracking along the \y -axis, and to solve the special case appearing in
region R1 for stations M2-M3.
Finally, the left-top and left-bottom processing units are connected to the right-bottom
and right-top processing units, respectively,
to exchange {\em corner} required by the muon tracking along the \y -axis.

Each processing unit is also connected to the backplane to insure data exchange
between boards.
This kind of exchange is performed via point to point links running
either at 40~MHz or at 1.6~Gbps.
\section{Implementation}\label{Implementation}
The implementation relies on high density FPGAs (Field Programable Gate Array) and
on massive use of multigigabit serial links deserialized inside the FPGAs.
Processors are connected to the outside world via optical links
while processing elements are interconnected with high speed
copper serial links.

The number of pins available on standard high density
connectors is not sufficient to transfer at 40 MHz the huge
amount of logical channels required to run the track finding
algorithm. Multiplexing the data at 80~MHz divides the
number of connections by a factor two, but the routing density
remains very high and therefore sensitive to cross talk. For this
reason, 1.6~Gbps serial links are mainly used for
the interconnection between processing modules.

By serializing most of the data exchanges at 1.6~Gbps,
the number of connections  is divided by a factor 16.
Sensitivity to cross-talk and to noise is decreased by a large
factor since links are routed on differential lines. However,
routing requires a lot of care since the geometry of the tracks
must be totally controlled to guarantee a good impedance
matching and to minimize electromagnetic emissions as well as sensitivity to
electromagnetic perturbations.

A processing board embeds 92 high speed serial links while the backplane assuring
the connectivity between the processing units uses mixed technologies:
288 single-ended links at 40~MHz and 110 differential serial links at 1.6~Gbps.
\subsection{The Processing Board}

\begin{figure*}
  \centering
  \includegraphics*[width=\textwidth]{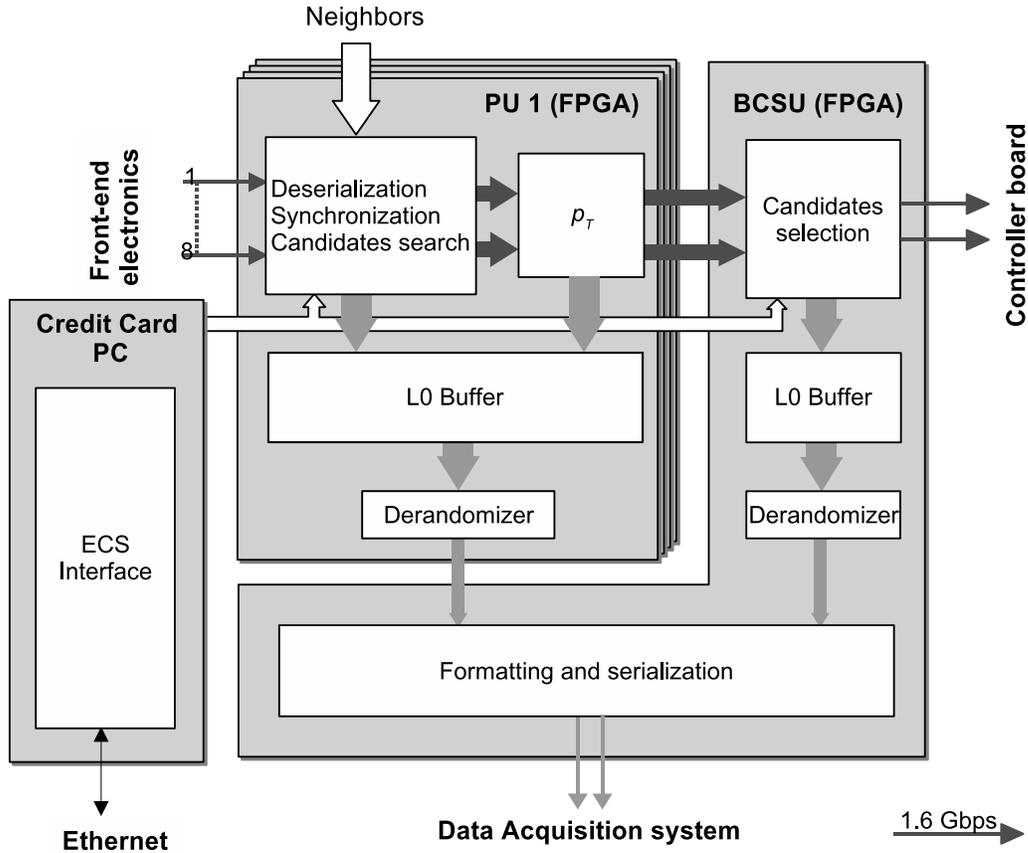}
  \caption{Hardware implementation of the processing board with four
           PUs, one BCSU. Interface with the experiment control system and the DAQ
           are also shown. Connections between PU and BCSU are only drawn for one PU
           for simplicity.}
\label{FIG_9}
\end{figure*}

The block diagram of the processing board is shown in Figure~\ref{FIG_9}.
Each processing element is implemented in an FPGA from the Stratix~GX family
embedding high speed serializers/deserializers.

The board sends data to the data acquisition system via a \lzero\ buffer/de\-ran\-do\-mi\-zer
housed in PUs and BCSU.
A \lzero\ buffer contains input and output data for a processing element.
Its size is equal to 544 bits for a PU and 352 bits for a  BCSU.
These buffers are managed by the BCSU which transfers their contents to the data
acquisition system via two high speed optical links at a maximum trigger rate of 1.1~MHz.

The interface to ECS (Experiment Control System) is based on an embedded PC
with a credit-card size~\cite{LHCbTDR}: SmartModule SM520PC produced
by Digital-Logic Inc. It is connected to the FPGAs by a local bus running at 20~MHz.
The credit-card PC downloads FPGA configurations and loads RAM as well as registers.
It is the main interface to control and debug a processor.

\begin{figure*}
  \centering
  \includegraphics*[width=\textwidth]{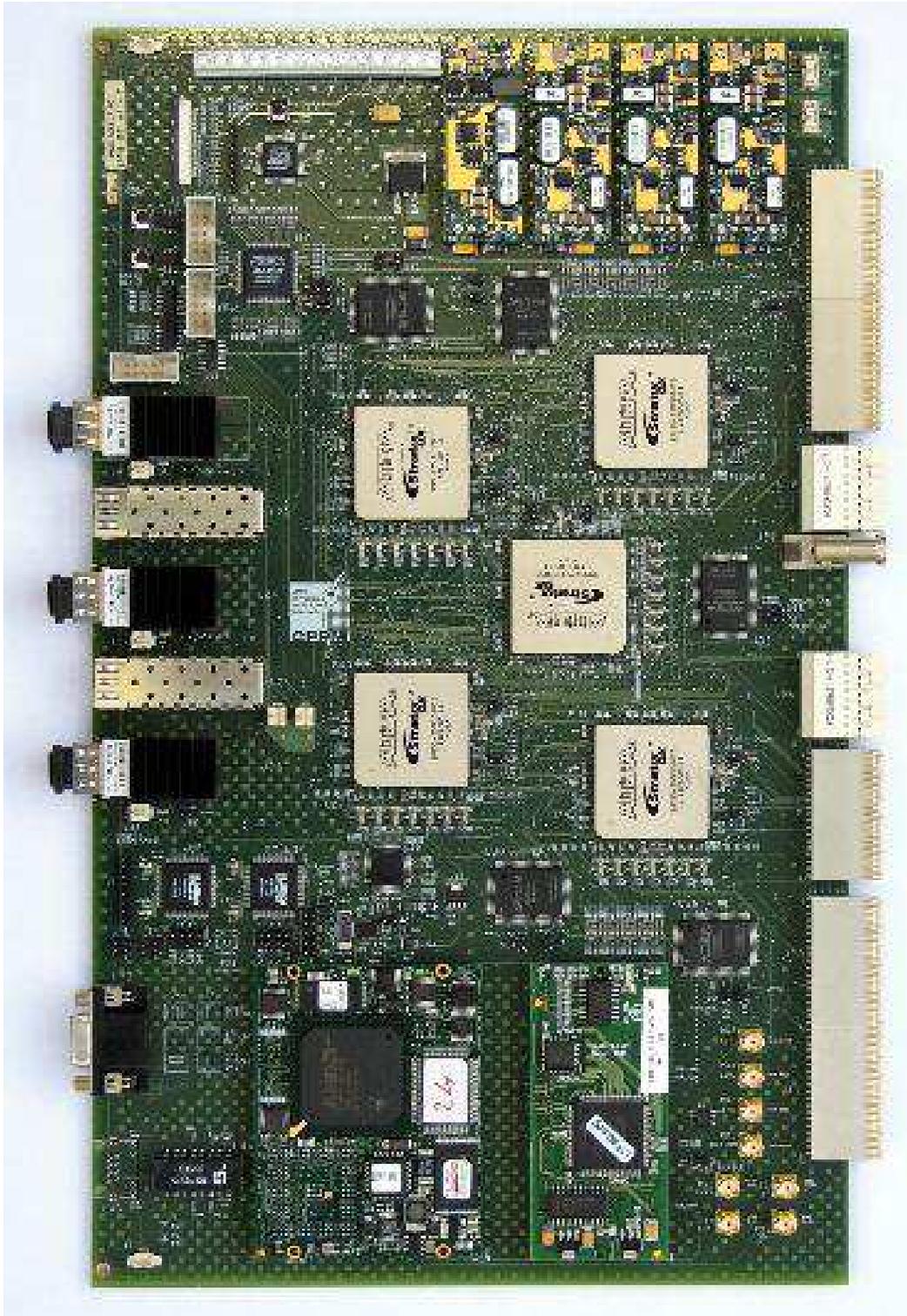}
  \caption{Photography of the processing board. The five FPGAs housing the four PUs
  and the BCSU are visible at the centre of the board. The three ribbon high speed
  transceivers are on the left side interleaved with two single emitters.
  DC/DC converters are on the top while the credit-card PC are on the bottom.}
\label{FIG_10}
\end{figure*}

The processing board is a 9U board shown in Figure~\ref{FIG_10}.
Details of its implementation are given in Appendix~\ref{AA1}.
\subsection{The Controller Board}

\begin{figure*}
  \centering
  \includegraphics*[width=\textwidth]{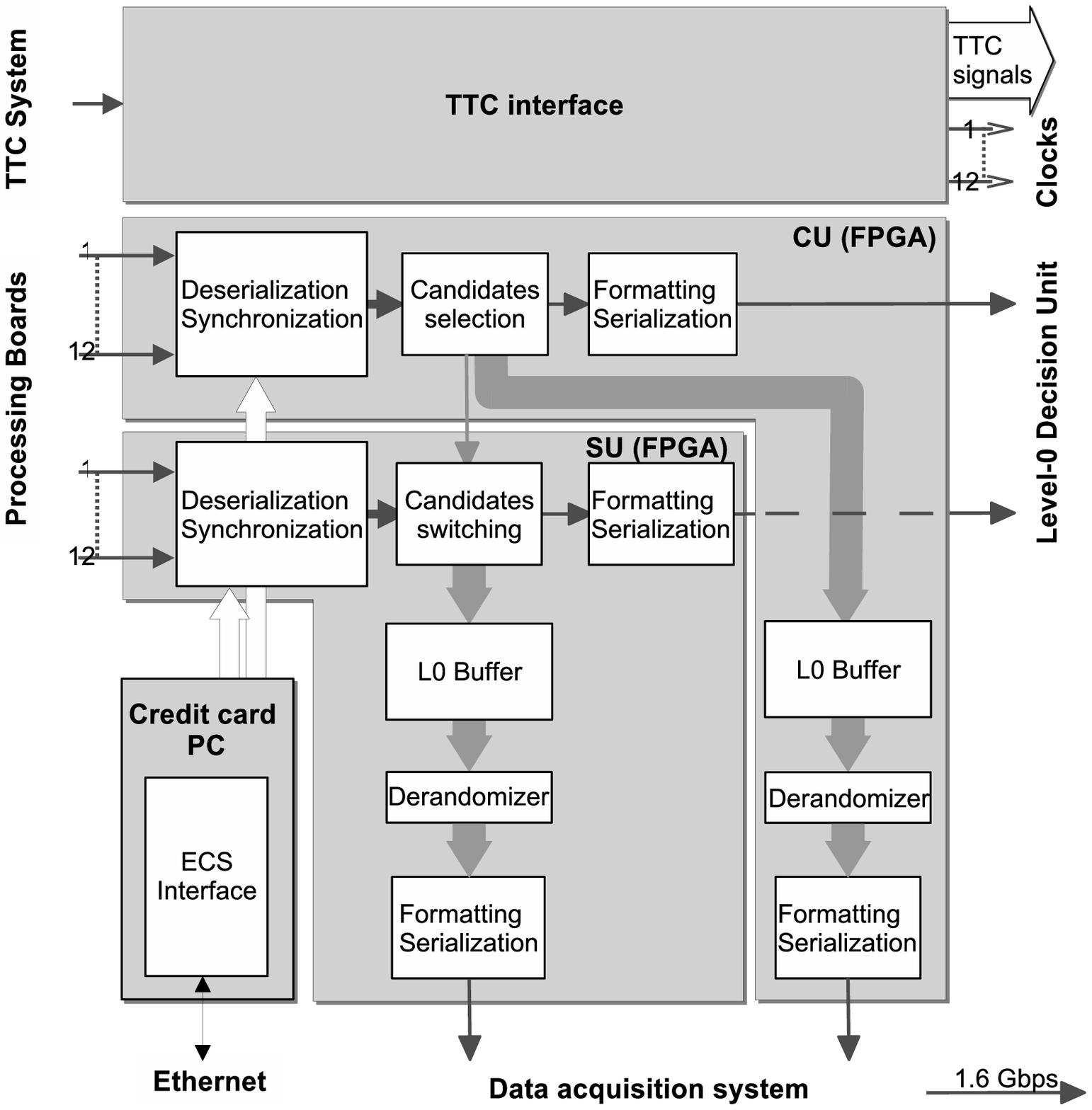}
  \caption{Hardware implementation of the controller board with its control and slave
           units. Interface with the TTC, \lzero\ Decision Unit, Experiment Control System and the DAQ
           are also shown.}
\label{FIG_11}
\end{figure*}

The block diagram of the controller board is shown in Figure~\ref{FIG_11}.
The board shares many common functionalities with the processing board:
credit-card PC,
\lzero -buffer/derandomizer, serializers/deserializers embedded in FPGAs and power distribution.
It contains two FPGAs from the Stratix~GX family since the number of
high speed deserializers embedded in a FPGA is limited to 16.
The first one is named {\em control unit}, the second one {\em slave unit}.

The controller board distributes the system clock and TTC signals~\cite{TTC} to the processing boards
through the backplane.
TTC information are received by an optical fiber and decoded by the TTCrx chip.
The 40~MHz clock is distributed using point to point connections while TTC signals
are broadcast using Gunning Transceiver Logic Plus (GTL+) standard.
This is a low voltage (1.5 V) technology with open drain output where
emitters conflicts are totally non-destructive for drivers.

Track candidates arrive from the backplane connector via 24 serial high speed links since
candidates information, coming from a processing board, is distributed over two serial links.
The first one contains the bunch crossing identifier, the transverse momenta and the M3
addresses of the candidates. The second one contains the candidate
addresses in M1-M2, as well as status and bunch crossing identification.
The first link is connected to the control unit, the second one to the slave unit.

The final candidates are sent to the \lzero\ decision unit via two high speed optical links.
One is driven by the control unit, the other to the slave unit.

The control unit and the  slave unit contain \lzero\ buffer and derandomizer
buffer. The size of the \lzero\ buffer is equal to 704 bits for a  control unit  and
720 bits for a slave unit.
The content of these two buffers is sent to the data acquisition system
via two high speed optical links at a maximum trigger rate of 1.1~MHz.

\begin{figure*}
  \centering
  \includegraphics*[width=\textwidth]{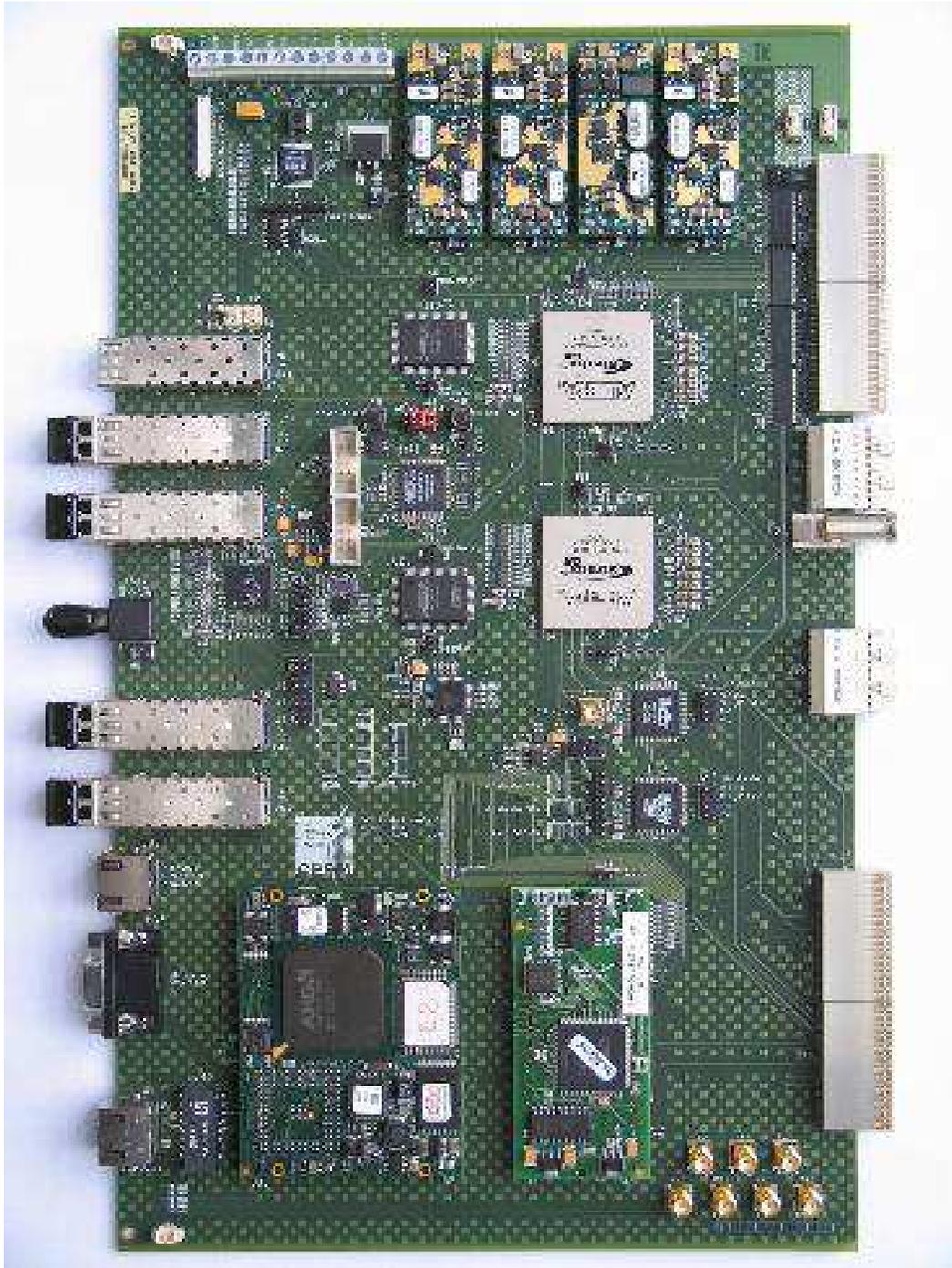}
  \caption{Photography of the controller board. The two FPGAs housing the control and
  the slave units are visible at the centre of the board. The four single high speed
  emitters are on the left side.
  DC/DC converters are on the top while the credit-card PC is on the bottom.}
\label{FIG_12}
\end{figure*}

The controller board is a 9U board, shown in Figure~\ref{FIG_12}.
Details of its implementation are given in Appendix~\ref{AA2}.
\subsection{The custom backplane}
The backplane contains 15 slots: twelve for the processing boards, one for the controller
board and two for test. The first test slot allows to check the processing board by looping its
outputs on its inputs. The second one allows to interface a logical analyser with a
processing/controller board.

\begin{figure*}
  \centering
  \includegraphics*[width=\textwidth]{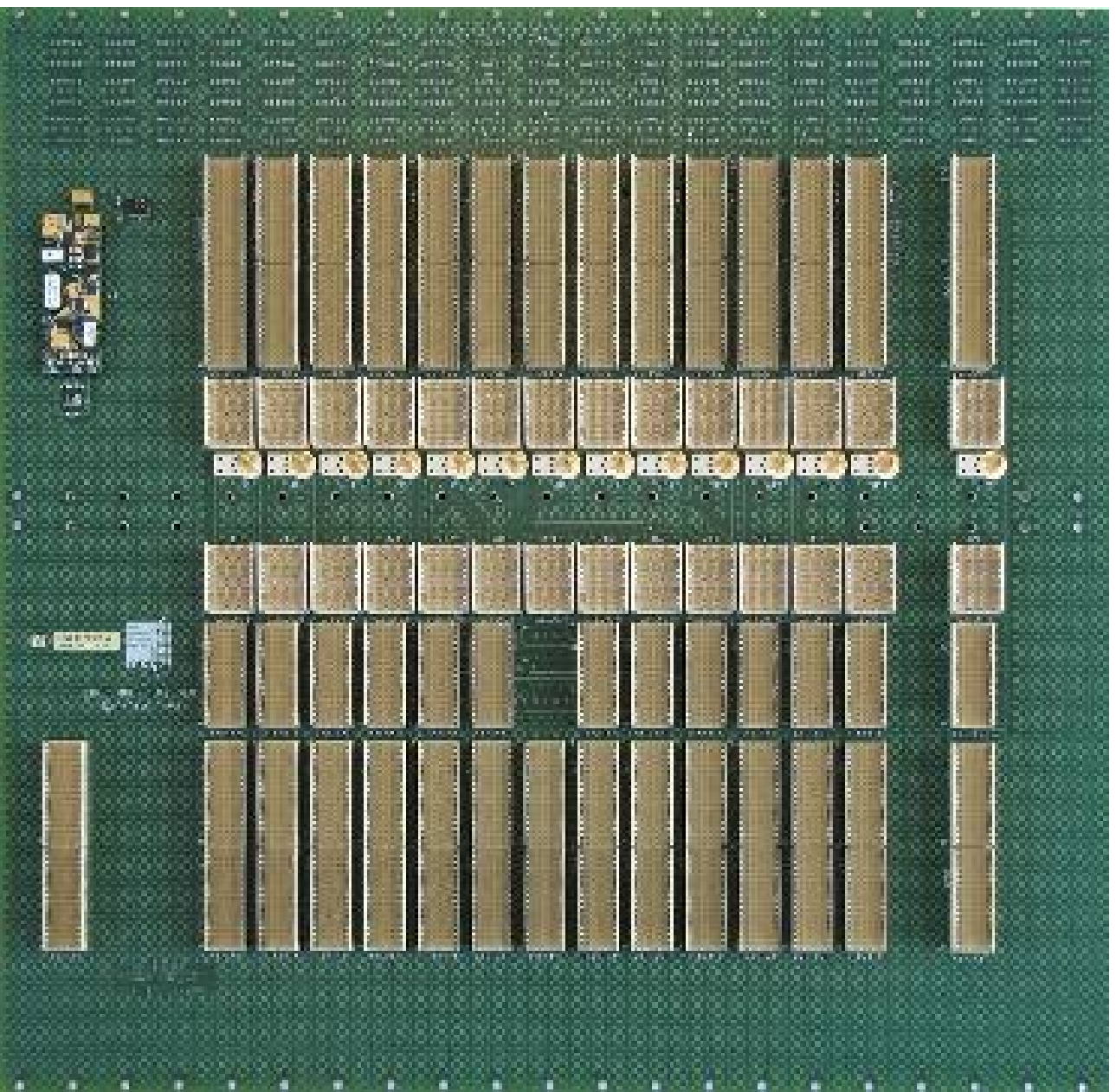}
  \caption{Photography of the backplane. The controller board is inserted
  in the central slot. The first row of connectors distributes power,
  system clock and TTC signals. The second and third rows are for the data exchange
  between processing boards via high speed serial links. The fourth row is for data
  exchange between PUs using point to point connections at 40~MHz.
  The last row is for test purposes.}
\label{FIG_13}
\end{figure*}

The backplane, shown in Figure~\ref{FIG_13}, distributes $+48$~V and $+5$~V power supplies,
ground, the 40~MHz clock and TTC control signals.
It permits the hit maps exchange between processing boards and the candidates collection
by the controller board.
Details of its implementation are given in Appendix~\ref{AA3}.
\subsection{Processing Unit}

\begin{figure*}
  \centering
  \includegraphics*[width=\textwidth]{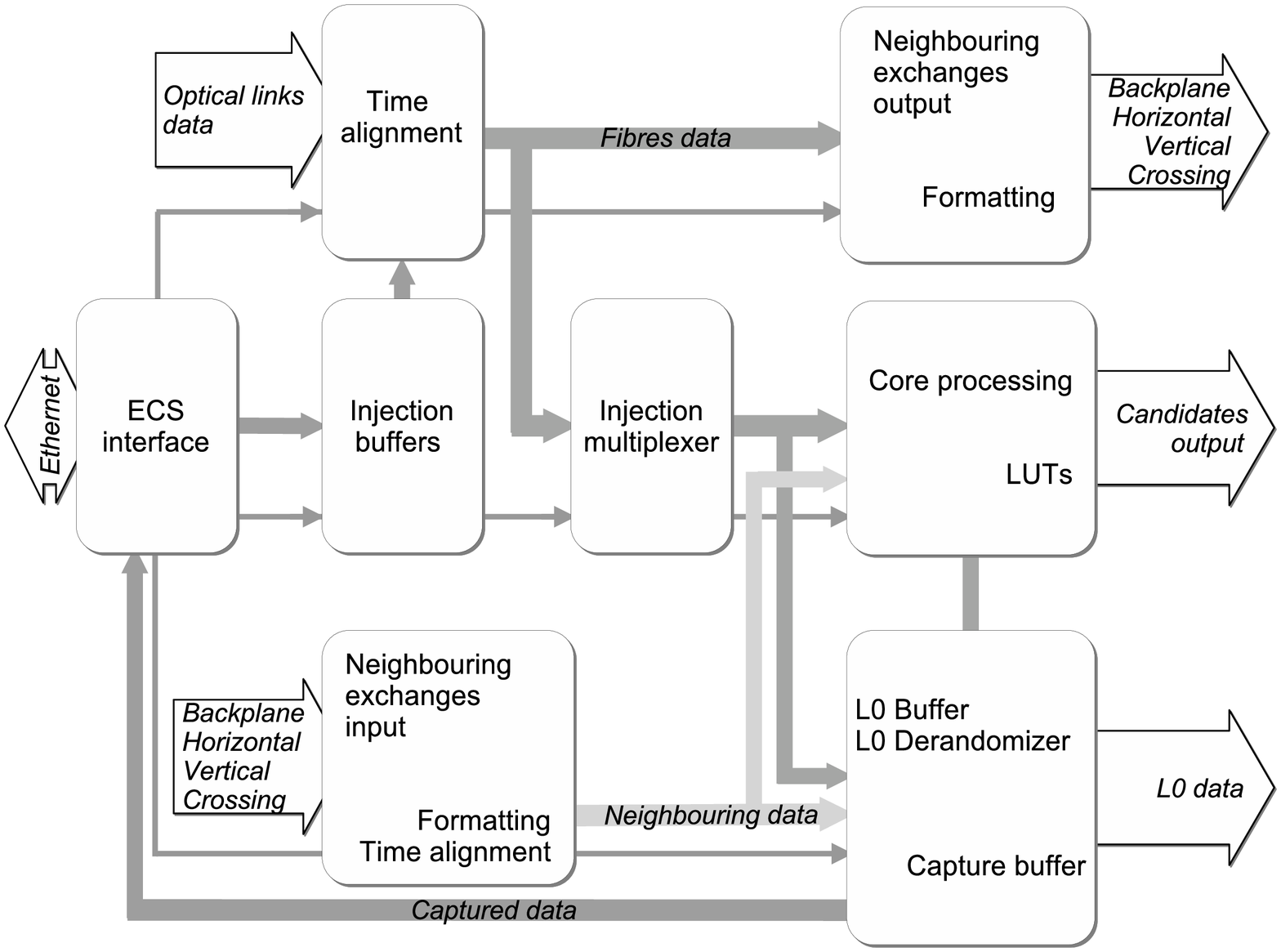}
  \caption{Block diagram of a processing unit.}
\label{FIG_14}
\end{figure*}

The block diagram for a PU is shown in Figure~\ref{FIG_14}.
It is subdivided in six main blocks:
\begin{enumerate}
  \item {\em Time alignment}\\
  Hit maps corresponding to a given bunch crossing arrive at different
  times at the outputs of optical links. They are time-and phase-aligned
  with the 40~MHz system clock using circular memories.
  \item {\em Neighbouring exchanges}\\
  The processing requires partial hit maps located in the neighbouring
  processing units. When the granularity of logical pads is not the same
  in the emitting and receiving PUs, a data formatting is performed.
  A time-alignment procedure is also necessary for neighbouring hit maps
  transported by serial links.
  \item {\em Core processing}\\
  When all pads/strips hit maps are available and phased with the system
  clock, they are transformed in pads.
  Local transformations are applied on pads with different granularity
  in order to run the track finding in an homogeneous space composed only
  of pads with the same granularity.
  Transverse momenta of candidates are computed using two look-up tables
  embedded in the FPGA.
  \item {\em \lzero\ pipeline buffer and derandomizer buffer}\\
  In order to wait for the \lzero\ trigger decision provided for each
  bunch crossing, data are stored for 105 consecutive bunch-crossings, in a buffer
  with a width of 532 bits and a maximum depth of 128.
  A 12-bit word containing the bunch-crossing identifier for
  the bunch-crossing accepted by the \lzero\ trigger is added.
  The resulting 544-bit word is stored in a derandomizer buffer with a depth of 16 words.
  The output of the derandomizer is transmitted to the BCSU as 34 words of 16-bit.
  \item {\em Injection buffers}\\
  These buffers are only used to debug a processing unit as well as a processing board.
  They mimic the input of the optical links for 16 consecutive events.
  Inputs and outputs of the processing are logged in the \lzero -buffer.
  In this test mode, they can be read by ECS through the \lzero\ derandomizer
  buffer for 16 consecutive events.
  \item {\em Capture buffer}\\
  During data taking, the ECS can not access the \lzero\ derandomizer buffer
  since it is actively used to send data to the data acquisition system.
  The capture buffer allows to make a snapshot of the
  \lzero\ buffers for one event accepted by the \lzero\ trigger.
  This functionality is one of the tools, developed to check the behavior of a
  processor during data taking.
\end{enumerate}
During the processing a 0--3563 {\em Bunch Crossing Identifier} travels with data
transported by all serial links.
In addition a {\em data valid} signal, tagging the start of
a new LHC cycle, is present on all point to point links running at 40-80~MHz or
1.6~Gbps. This information guarantees the time alignment during the processing.

Details of the implementation for a PU can be found in Appendix~\ref{AA4}.
\section{Technologies}\label{Technologies}
In this section, more details on the technologies used for
the \lzero\ muon trigger are given.
\subsection{High speed optical links}\label{OL}
High speed serial transmission reduces the number of signal lines required
to transmit data from one point to another. It also offers a high level of integration
with many advantages:
high reliability for data transfer over 100~meters;
complete electrical isolation avoiding ground loops and common mode problems.
In addition, the integration of several high speed optical links in a single device
increases data rate while keeping the component count manageable at a reasonable cost.

Ribbon optical links integrate twelve optical transmitters (fibres, receivers) in one module.
The important benefit of ribbon optical links is based on
low-cost array integration of electronic and opto-electronic components. It also
provides a low power consumption and a high level of integration.

\begin{figure*}
  \centering
  \includegraphics*[width=\textwidth]{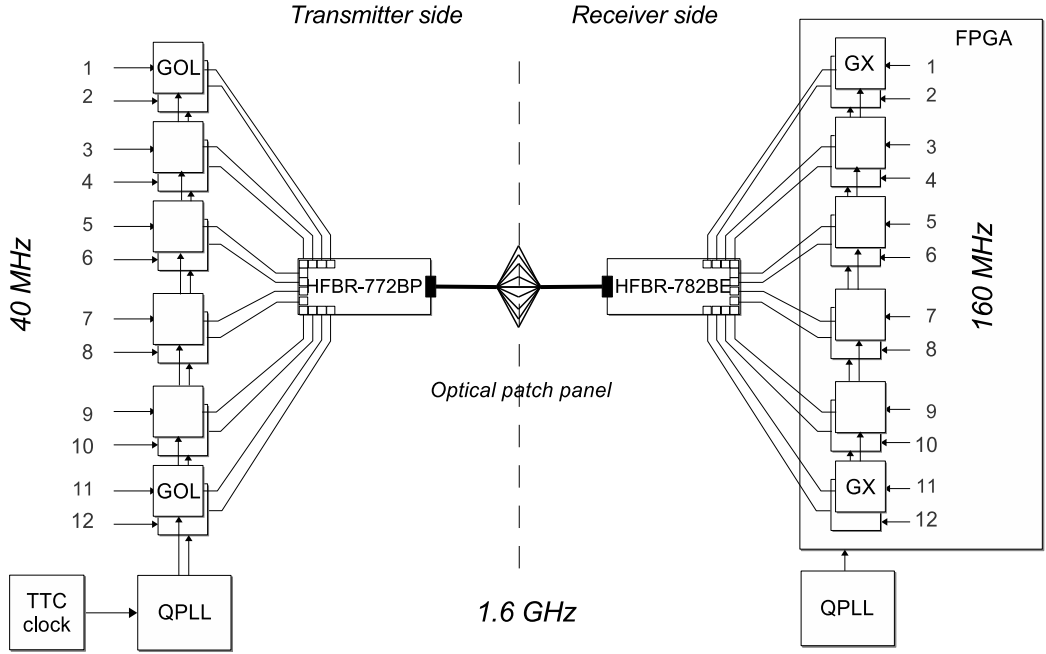}
  \caption{Overview of the ribbon optical link.}
\label{FIG_15}
\end{figure*}

An overview of ribbon optical link developed for the \lzero\ trigger is shown in Figure~\ref{FIG_15}.
The emitter stage relies on twelve serializer chips connected to one optical transmitter.
The serializer is the GOL~\cite{GOL}, a radiation hard chip designed by the CERN microelectronic group,
transforming every 25~ns 32-bit words into a serial signal cadenced at 1.6~Gbps using a {\em 8B/10B} encoding.
High frequency signals are converted into an optical signal by the 12-channels optical
transmitter from Agilent HFBR-772BE.
The module is designed to operate with multimode fibers at a nominal wavelength of 850~nm.

Initially, the LHC clock distribution was not intended to be used for optical data transmission.
Hence, it does not fulfill the severe jitter constraints required by high speed serializers.
The GOL requires a maximum jitter of 100~ps peak to peak to operate correctly whereas
the LHC clock jitter is as large as 400 or 500~ps.
To reduce the jitter, a radiation hard chip, the QPLL~\cite{QPLL}, was designed by the CERN microelectronics group.
It reduces the jitter to an acceptable value with the help of a reference quartz crystal
associated with a phase locked loop.

The emitter side is close to the detector in a place where SEUs are expected.
The GOL and QPLL chips are radiation hard chips immune to SEU.
However, the optical transceiver is a commercial component designed to work in an environment
free of radiation. An irradiation campaign took place at the Paul Scherrer Institute in December 2003.
The optical transceiver works within its specification up to a total dose of ~150~Gy. The cross-section
for single event upsets is equal to $(4.1 \pm 0.1)\times 10^{-10}$~cm$^2$ per single optical link.

The connections between the front-end electronic boards and the processing board consist
of ribbons of twelve fibers with MPO (Multi-fiber Push-On) type connectors on both side ($\sim10$~m.),
MPO-MPO patch panels, long cables containing eight ribbons with MPO connectors ($\sim80$~m.),
fanout panels (MPO-MPO or MPO-SC), short ribbons of twelve fibers ($\sim3$~m) with a MPO
connector on one side and a MPO or 12 SC connectors on the other side.

The receiving side is the mirror of the emitting side.
Optical signals are converted into 1.6~Gbps serial electrical signals by the 12-channels optical receiver
HFBR-782BE. The twelve high frequency signals are deserialized by the GX block embedded in Stratix~GX FPGA's.

The transmission used by the GX blocks is based on the physical layer of the IEEE802.3 standard.
Data transmitted are encoded according to the 8B/10B scheme.
The 8B/10B scheme encodes 8-bit words into 10-bit words.
The 10-bit codes are chosen to contain: either 6 zeros and 4 ones; or 4 zeros and 6 ones;
or 5 zeros and 5 ones.
The purpose of this encoding is to balance the number of ones and zeros transmitted
and therefore to avoid the appearance of a DC level.
Each 8-bit data word has two complementary representations in the 10-bit space:
a positive and a negative one.
The representation that balances the quantity of zeros
and ones is chosen for each word and transmitted.
This protocol is entirely managed by the GX buffers.

The size of data words is equal to 32 bits, their frequency is 40~MHz.
Each data word, is multiplexed on an 8-bit path at 160 MHz to reach the GX block.
The latter serialize the data on a differential path at 1.6 Gbps.
The reception side is symmetrical: data are demultiplexed by the GX blocks on an
8-bit path at 160~MHz and an internal logic demultiplexes the data on a 32-bit path at 40~MHz.
Moreover for each channel a reception clock is extracted from the data by the GX blocks.
All the reception clocks are not necessarily in phase.
Data are aligned in time with the global system clock by a dual port memory mechanism implemented
with the help of internal cells of the FPGA.

Performance of the optical link have been measured with several setups in different ways.
The bit error rate measured with Lecroy SDA11000 Serial Data Analyser is below $10^{-16}$
for a single 100 meter long fiber.
\subsection{Debugging features of Field Programable Gate Arrays}
The growth in size and complexity makes design verification a critical bottleneck
for  FPGA systems.
To help with the process of design debugging, Altera provides the possibility to build a
logic analyzer in the unused cells and memory of a FPGA.
We heavily rely on this feature to examine the behavior of internal signals,
without using extra I/O pins, while the algorithm is running at full speed.
Sampling clock and custom trigger-condition logic are defined.
All captured signal data are stored in device memory until the designer is ready
to read and analyze the data.
Then they are transmitted by a JTAG serial link to a control PC and displayed on
screen with a logic-analyzer-like look.
\subsection{The layout of printed circuit boards}
Because of the density of the design, 18 layers are required to route the signals
on a processing board.
The stack-up uses a {\em power plane / signal / signal / power plane} topology which
reduces the number of layers required to route the signals and at the same time
allows to control their impedance.
To match the internal impedance of the GX blocks the individual impedance of the
tracks is set to 55~$\Omega$ and for differential pairs to 107~$\Omega$.

Track width and track separation have been limited to 120~$\mu$m to ease the manufacturing.
The high speed signals are referenced to analog ground and power
to decrease the noise induced by hundreds of signals switching synchronously at 40~MHz.
Analog and digital grounds are completely independent on the board.
They are merged on one point in the backplane.

The controller board and the backplane being slightly less complex, they respectively
contain 14 and 18 layers based on a more classical {\em power plane / signal / power plane} stack-up.
All the other features are identical to those of the processing board.

A purely manual routing is impossible due to the large number of signals and to the high density.
On the other hand a completely automatic routing leads to topologies that
do not work because they generate reflections destroying the signal integrity.
To solve this issue, an iterative approach in three steps have been used:
\begin{enumerate}
  \item simulate for each kind of driver the most appropriate topology and
  derive constraints that are given to the automatic router;
  \item run the automatic router;
  \item validate the result by running post-routing simulation to check the
  signal integrity.
\end{enumerate}
In this way, the automatic router produces a correct routing for a large fraction of the
the full design. As an example, for the processing board, the router has to deal with an
extremely high number of nets, 10~000.
It failed only for 150 nets which were routed manually.
\subsection{Quality assurance during manufacturing}
During the PCB manufacturing, two kinds of tests are performed:
\begin{enumerate}
  \item check electrical connections to guarantee that the connectivity
  is absolutely identical to the CAD design.
  \item measure characteristic impedance for reference tracks. Keep only boards where
  impedance match specification for all layers.
\end{enumerate}

During the assembly process, four types of test are performed:
\begin{enumerate}
  \item Connections are verified using X-Rays on all FPGAs. Since FPGAs use
  Ball Grid Arrays packaging connections cannot be visually checked after soldering.
  \item Boards are heated from 0 to 70~$^\circ$C in two cycles of approximately 30~hours.
  The role of this test is to mechanically constrain the board to accelerate the
  appearence of early life failures.
  In this way, failures are detected directly at the manufacturer level and the
  reliability of the boards is improved.
  \item A flying probe control is made on the board after soldering.
  This test allows to detect shorts or open circuits, to control the value of
  passive components and eventually their polarity.
  The test is made by direct contact on the component pins or connectors
  or through the housing by Hall effect for plastic BGAs.
  The test covered 94.3\% of the components for a processing board.
  Unfortunately the five FPGAs of the processing board are among the 5.7\%
  left out by this test.
  The metallic shielding of the housing does not allow any measurement by Hall effect.
  As a consequence, only 26\% of the connections are covered by the flying probe test.
  \item A boundary scan test (JTAG) was implemented.
\end{enumerate}
\section{Debugging and Monitoring tools}\label{Debug}
Each board embeds a credit-card PC, running Linux, interfaced to FPGAs by a custom 16-bit bus.
By that way, the operation of any FPGA of the system is controlled and monitored
through error detection mechanisms, error counters, spy and snooping mechanisms.

The \lzero\ muon trigger is a very complicated system.
Any malfunction can therefore be difficult to understand and to interpret.
We developed the {\em interconnection matrix test}, to verify exhaustively the connectivity
of the system and the {\em functional test} relying on a software emulator.

The emulator and \lzero\ buffers are the key components to debug the system
and to validate its functionality, at any time, during a data taking period
or later on in the ultimate phase of data analysis.
\subsection{Emulator of the \lzero\ muon trigger}
Each processing element logs its inputs and results of its computation in a \lzero\ buffer.
The width of \lzero\ buffers varies between 352 and 720-bits resulting to an event size of
about 4~kBytes for a processor.
Zero suppression and data compression algorithms applied later on in the data acquisition chain,
reduce the event size below 0.3~kBytes.

To understand such a large quantity of information, we developed the emulator software
reproducing the behaviour of the hardware on a bit to bit basis.
By comparing \lzero\ buffer contents with those produce by the emulator run on the same input data,
we can isolate any error and understand what has happenned.

The emulator is implemented in C++ and is based on two generic classes:
the {\em unit} and the {\em register}.
A register is a bank of data identified by a name, centrally controlled by a register factory.
A unit is a simple object containing input and output registers and
which can execute a function on them. It may also contain other units
and trigger the execution of their function.

To simulate the \lzero\ muon trigger, units are specialized to
represent a processor component such as a processing board or a
processing element. The emulator is configured using the database that
describes the processor topologies. It then forms a hierarchical system
of units that communicates through a set of formatted registers
reproducing exactly the data transferred in the processor.
\subsection{Test of interconnection matrix}
The interconnection matrix between FPGAs is complicated and depends on the processing board.
To validate it globally, we developed a dedicated software called {\em Spyd}.

All FPGAs of a \lzero\ muon processor are loaded with a unique firmware which
can validate all links running at 40~MHz, 80~MHz and 1.6~Gbps in parallel.
Each interconnection has an emitter and a receiver side. They are configure differently.

On the emission side, a frame of 2048 words is emitted continuously, one word every 25~ns.
The first 12 words of the frame form the header. For each of them, the header tag is encoded
on the three LSB bits\footnote{Encoding the header on three LSB bits guarantee its decoding
for all buses of the system since their width varies between 3 and 54 bits.}.
Four words are used to synchronize the emitter with the receiver.
The remaining eight words provide the address of the emitter: slot in the crate,
FPGA on the board and port number in the FPGA.
Data words of the frame merge several 6-bit counters to fill the width of the link.

On the reception side, the behaviour of a link is checked by comparing
received words with expected ones, every 25~ns.
The receiver logs two types of errors in dedicated registers: no-synchronization and
words error count on 16-bit. It also keeps the address of the emitter.

When this test is running for a complete processor crate, 570 serial links and 289 point-to-point
connections, at 40 or 80~MHz, are running in parallel.
We developed a dedicated software based on a client server protocol to monitor the test.

The processing board is generic: hence all possible connections with the backplane are available.
However, on our custom backplane only relevant connections are implemented.
In such configuration, some links are always in error since an emitter is not connected to a receiver
and missing links have to be removed from the error analyzer
using the topology database.

The software is a distributed application composed of a master process running on a supervision
station and slave processes running on credit-card PCs.
At the end of the test period, the application collects link status and error counters to produce
a full error report.
This client-server application is written in python using a socket server module and XML messages.

The Spyd test was run on the first muon processor during 26 hours without any errors detected !
\subsection{The functional test}
A {\em functional test} validates the functionality of a \lzero\ muon trigger processor covering
the VHDL programming of the track finding as well as the internal time-alignment mechanism.

The \lzero\ muon trigger emulator is used in the simulation of the LHCb experiment.
Hit maps for optical links and simulated \lzero\
buffers contents are extracted from Monte-Carlo events.

All FPGAs of a muon processor are loaded with the \lzero\ muon trigger configurations.
Hit maps are pushed in the injections buffers per block of 16 consecutive events.
The processor crunches them at the nominal speed.
The contents of the \lzero\ buffers are read and automatically compared with those from the emulator.

A dedicated software based on a client-server approach has been developed.
This test is usually run on $10^4$ events and takes about ten minutes.
Most of the time is spent in writting injection buffers and reading \lzero\ buffers.
\section{Conclusions}\label{Conclusions}

\begin{figure*}
  \centering
  \includegraphics*[width=\textwidth]{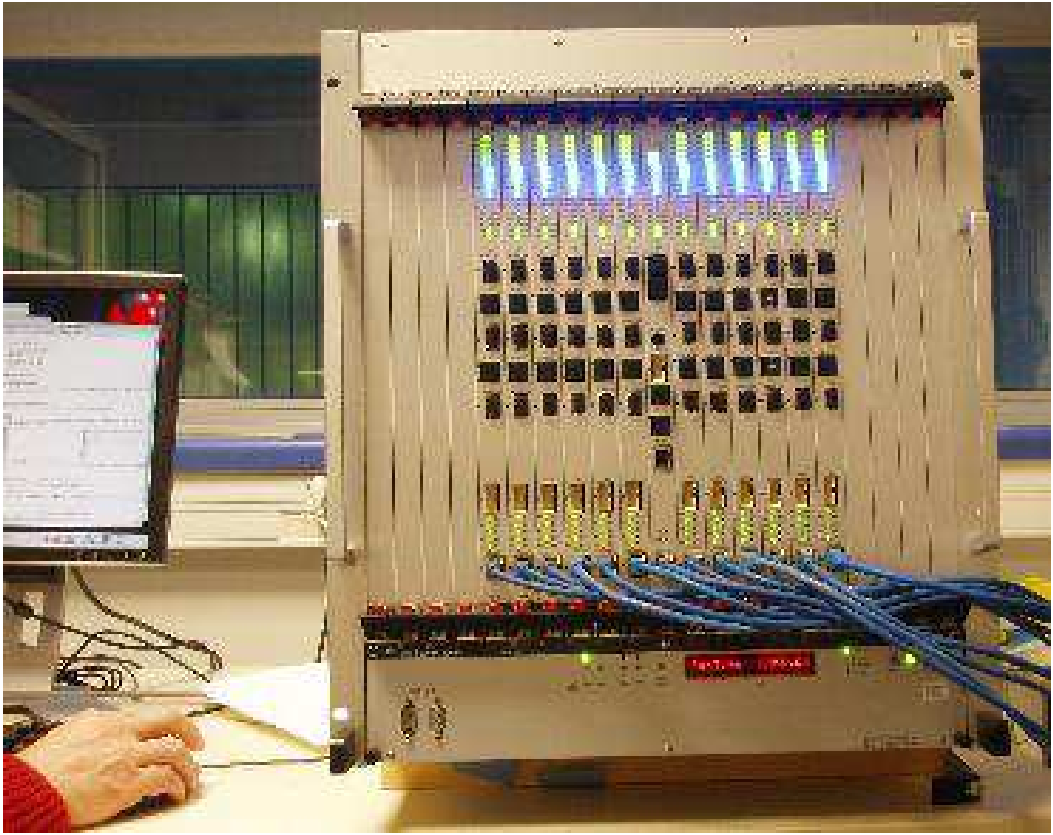}
  \caption{Photography of the first \lzero\ muon processor in December 2006.}
\label{FIG_16}
\end{figure*}

The \lzero\ muon trigger requires a very innovative architecture to handle the
complex data flow, the large volume of data and the high input rate.
It relies on a large number of: high speed optical links, high density FPGAs and
high speed serial links between FPGAs.
The complex data flow has been described by a dedicated software tool which was used
at each step of the design and qualification.
This software guarantees a perfect correspondance between the hardware and the trigger emulation
used in the LHCb simulation. It is also the key component to understand rapidly
any possible malfunctions at any time.

The architecture of the \lzero\ muon trigger was finalized in September 2004.
Sixteen months later, the first processor, shown in Figure~\ref{FIG_16}, was produced and tested.
The commissioning in the LHCb experiment started in February 2007 and will end during the summer of 2007.

\section*{Acknowledgements}
We would like to thank the CERN LHCb staff for their support and co-operation, as well as
the technical and engineering staff of our institute.
This work was supported by the French CNRS/Institut National de Physique Nucléaire et de Physique
des Particules and the Universit\'e de la M\'editerran\'ee.

%
\appendix
\section{Appendices}\label{AA}
\subsection{The processing board}\label{AA1}
The processing board is a 9U board with a width of 220~mm, a thickness of 2.7~mm and a number of layers equal to 18.
The minimal size of a track and the minimal separation between two tracks is 120~$\mu$m.
The number of differential high speed serial lines is 92.
They are routed on dedicated layers. Their impedance is controlled and should match $107\pm11$~$\Omega$.
We use the same type of FPGA for the processing units and the best candidates selection unit. It
is a Stratix~GX EP1S25FF1020-C6 from Altera with 16 SERDES running from
622 Mbps up to 3.125 Gbps. A FPGA is connected to the printed circuit via
1020 pins with a pitch of 1~mm.
A total of 1532 components are mounted on the board with
seven pressfit connectors allowing connections to the backplane.
Two of them convey high speed serial data. We use
ERmet ZD Hard Metric high speed connector from ERNI
embedding 30 pairs of signal with internal differential shielding
and a differential impedance around $100$~$\Omega$.

The board is powered through the backplane by two independent sources: $+48$~V
and $+5$~V. DC/DC converters embedded on the board generate $+1.5$, $+3.3$~V from the
$+48$~V. The QPLL chip (see section~\ref{OL} ) requires $+2.5$~V.
It is obtained from $+3.3$~V using a regulator.
The power consumption of the board is below 60~W.
\subsection{The controller board}\label{AA2}
The controller board shares a lot of functionality with the processing board.
Therefore their implementation are very similar.

The controller board is a 9U board with a width of 220~mm,
a thickness of 2.6~mm and a number of layers equal to 14.
The minimal size of track and the minimal separation between two tracks is 120~$\mu$m.
The number of differential high speed serial links is 28.
They are routed on dedicated layers.
Their impedance is controlled and should match $107\pm11$~$\Omega$.
We use the same FPGA for processing and controller boards.
In addition, 1032 components are mounted on the board with six pressfit connectors.
Two of them are ERmet ZD Hard Metric high speed connectors from ERNI.
\subsection{The custom backplane}\label{AA3}
Dimensions of the backplane are $426.72\times395.4$~mm with a thickness of 3.4~mm.
It is an eighteen layers printed circuit. The minimal size of tracks and the minimal
separation between two tracks are 120~$\mu$m.
The number of pressfit connectors is 97.
The number of differential high speed serial lines is 110.
They are routed on dedicated layers.
Their impedance is controlled and should match $107\pm11$~$\Omega$.

The number of point to point data lines running at 40~MHz is 288.
Point to point traces for clock signals were tuned in order to obtain the same
propagation delay for all processing boards.
The GTL+ standard used for the broadcast control signals needs pull-up resistors
to $+1.5$~V.
The $+1.5$~V is achieved from the $+48$~V, using DC/DC converter mounted on the backplane.
\subsection{The processing Unit}\label{AA4}
The processing unit is implemented in Stratix GX EP1SGX25FF1020C6 from Altera
using 47\% of logic elements, 32\% of memory bits, 5 of the 8 PLLs,
all high speed deserializers, half of high speed serializers,
none of the DSP blocks and 65\% of I/O pins.

\newpage\listoffigures
\end{document}